\documentclass[pre,twocolumn,10pt,aps]{revtex4-1}

 \usepackage{bm}
 \usepackage{verbatim}
 \usepackage{amsmath}
 \usepackage{amssymb}
 \usepackage{amsthm}
 \usepackage{latexsym}
 \usepackage{amsfonts}
 \usepackage{epsfig}
 \usepackage{epstopdf}
 \usepackage{wasysym} % for certain special symbols
 \usepackage{subfigure}
 \usepackage{color}
 \definecolor{darkblue}{rgb}{0,0,.5}
 \usepackage[linktocpage, colorlinks=true, linkcolor=darkblue, citecolor=darkblue]{hyperref}
 \usepackage[all]{hypcap}
 \usepackage[makeroom]{cancel}
 
\newcommand{\C}[1]{{\cal{#1}}}
\newcommand{\bb}[1]{\textbf{#1}}

\begin{document}

\title{Autonomous implementation of thermodynamic cycles at the nanoscale}

\author{Philipp Strasberg$^1$}
\author{Christopher W. W\"achtler$^{2,3}$}
\author{Gernot Schaller$^{2,4}$}
%\author{Mar\'ia Garc\'ia D\'iaz}
%\author{Andreas Winter$^{1,2}$}
%\email{philipp.strasberg@uab.cat}
\affiliation{$^1$F\'isica Te\`orica: Informaci\'o i Fen\`omens Qu\`antics, Departament de F\'isica, Universitat Aut\`onoma de Barcelona, 08193 Bellaterra (Barcelona), Spain}
\affiliation{$^2$Institut f\"ur Theoretische Physik, Sekr. EW 7-1, Technische Universit\"at Berlin, Hardenbergstr. 36, 10623 Berlin, Germany}
\affiliation{$^3$Max Planck Institut f\"ur Physik komplexer Systeme, N\"othnitzer Str. 38, D-01187 Dresden, Germany}
\affiliation{$^4$Helmholtz-Zentrum Dresden-Rossendorf, Bautzner Landstra\ss e 400, 01328 Dresden, Germany}
%\affiliation{$^2$ICREA -- Instituci\'o Catalana de Recerca i Estudis Avan\c{c}ats, Pg.~Lluis Companys, 23, 08010 Barcelona, Spain}

\date{\today}

\begin{abstract}
 There are two paradigms to study nanoscale engines in stochastic and quantum thermodynamics. Autonomous models, which 
 do not rely on any external time-dependence, and models that make use of time-dependent control fields, often combined 
 with dividing the control protocol into idealized strokes of a thermodynamic cycle. While the latter paradigm offers 
 theoretical simplifications, its utility in practice has been questioned due to the involved approximations. Here, we 
 bridge the two paradigms by constructing an autonomous model, which implements a thermodynamic cycle in a certain 
 parameter regime. This effect is made possible by self-oscillations, realized in our model by the well studied 
 electron shuttling mechanism. Based on experimentally realistic values, we find that a thermodynamic cycle analysis 
 for a single-electron working fluid is {\it not} justified, but a few-electron working fluid could suffice to 
 justify it. Furthermore, additional open challenges remain to autonomously implement the more studied Carnot and 
 Otto cycles. 
\end{abstract}

\maketitle
%\tableofcontents

%\newtheorem{mydef}{Definition}[section]
\newtheorem{lemma}{Lemma}[section]
%\newtheorem{thm}{Theorem}[section]
%\newtheorem{crllr}{Corollary}[section]
%\newtheorem*{thm*}{Theorem}%[section]
%\theoremstyle{remark}
%\newtheorem{rmrk}{Remark}[section]

%%%%%%%%%%%%%%%%%%%%%%%%%%%%%%%%%%%%%%%%%%%%%%%%%%%%%%%%%%%%%%%%%%%%%%%%%%%%%%%%%%%%%%%%%%%%%%%%%%%%%%%%%%%%%%%%%%%%%%%%
\emph{Introduction.---}The success of thermodynamics builds on the possibility to reduce macroscopic phenomena to a 
few essential elements. An important role in that respect has played the idea of a thermodynamic \emph{cycle}, 
allowing to break up the working mechanism of a complex machine into steps, which are easy to study. 
These steps are called, e.g., adiabatic, isothermal or isentropic \emph{strokes}. 

Understanding thermodynamics at the nanoscale forces us to give up many traditionally used assumptions. From that 
perspective, it is interesting to observe that much current work focuses on idealized cycles as introduced by, e.g., 
Carnot and Otto back in the 19th century; see Refs.~\cite{VinjanampathyAndersCP2016, KosloffRezekEntropy2017, 
GhoshEtAlBook2018, FeldmannPalaoBook2018, LevyGelbwaserKlimovskyBook2018, DeffnerCampbellBook2019} for reviews. But 
for a small system, such 19th-century-cycles seem to be based on crude assumptions: the system needs to be repeatedly 
(de)coupled from a bath and work extraction is modeled semi-classically via time-dependent fields. 

Recent experiments implementing thermodynamic cycles in nanoscale engines echo these 
problems~\cite{BlickleBechingerNatPhys2015, MartinezEtAlNP2016, RossnagelEtAlScience2016, PassosEtAlPRA2019, 
KlatzowEtAlPRL2019, VonLindenfelsEtAlPRL2019, PetersonEtAlPRL2019}: the thermal baths are typically \emph{simulated} 
via additional time-dependent fields and a \emph{net} work extraction (including the work spent to generate the 
driving fields) has not been demonstrated. This has raised doubts about the usefulness of cycles to analyze nanoscale 
engines (see the recent discussion~\cite{QuoVadis}). Yet, a critical theoretical study to rigorously address 
this problem is missing. 

Here, we provide such a critical study based on the phenomenon of \emph{self-oscillations}~\cite{JenkinsPR2013}. 
This provides a missing link between nanoscale engines studied with a cycle analysis and \emph{autonomous} engines, 
such as thermoelectric devices~\cite{SchallerBook2014, SothmannSanchezJordanNanotechnology2015, BenentiEtAlPhysRep2017, 
WhitneySanchezSplettstoesserBook2018} or absorption refrigerators~\cite{MitchisonCP2019}. To be precise, by 
``implementing a thermodynamic cycle autonomously'' we mean that (see Refs.~\cite{TonnerMahlerPRE2005, DeffnerJarzynskiPRX2013, GelbwaserKlimovskyKurizkiPRE2014, MayrhoferEtAlPRB2021} for related ideas): 
\begin{enumerate}
 \item[\emph{(i)}] The starting point is a model \emph{without} explicit time-dependence. The guiding principle should 
 be simplicity and experimental feasibility. 
 \item[\emph{(ii)}] In some parameter regime the \emph{dynamics} of the model reduces to that of a thermodynamic cycle. 
 \item[\emph{(iii)}] For a subset of the parameter regime in \emph{(ii)}, the \emph{thermodynamics} of the cycle 
 analysis matches the original thermodynamics of the autonomous model. 
\end{enumerate}

In particular, by using experimentally realistic values, 
we can draw practically relevant conclusions at the end. Moreover, similar to autonomous Maxwell 
demons~\cite{StrasbergEtAlPRL2013, HartichBaratoSeifertJSM2014, HorowitzEspositoPRX2014, KoskiEtAlPRL2015, 
StrasbergEtAlPRB2018, PtaszynskiEspositoPRL2019, SanchezSamuelssonPottsPRR2019, SanchezSplettstoesserWhitneyPRL2019}, 
our work bridges a gap between different theoretical paradigms, as well as between theory and experiment. 

\begin{figure}%[b]
 \centering\includegraphics[width=0.30\textwidth,clip=true]{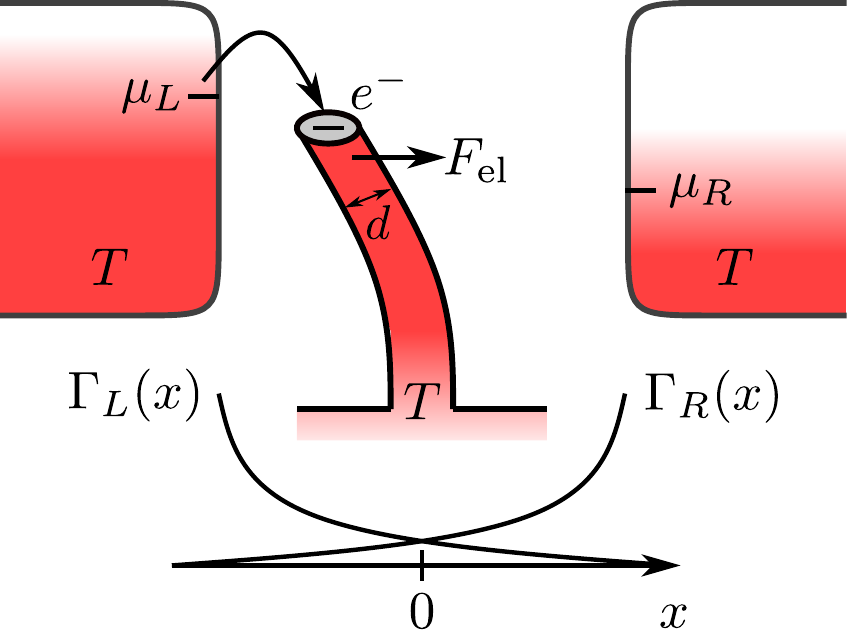}
 \label{fig sketch} 
 \caption{Two leads with chemical potentials $\mu_L$ and $\mu_R$ at temperature $T$ are placed at some distance. In 
 between, a quantum dot (grey disk) is mounted on a nanopillar, which in turn is mounted on a larger solid at 
 temperature $T$. The nanopillar has diameter $d$ (which is relevant for our final discussion) and can oscillate from 
 left ($x<0$) to right($x>0$). The left/right bare tunneling rates $\Gamma_{L/R}(x)$ depend on the dot position as 
 sketched at the bottom. If an electron jumps on the dot, 
 an electrostatic force $F_\text{el}$ acts on the nanopillar towards the right. The sketch pictures experimental 
 setups of Refs.~\cite{ErbeEtAlPRL2001, ScheibleBlickAPL2004, KimQinBlickNJP2010, KimParkBlickPRL2010, 
 KimPradaBlickACS2012}, but the theoretical description below holds for a wider range of self-oscillating nanosystems. }
\end{figure}

%%%%%%%%%%%%%%%%%%%%%%%%%%%%%%%%%%%%%%%%%%%%%%%%%%%%%%%%%%%%%%%%%%%%%%%%%%%%%%%%%%%%%%%%%%%%%%%%%%%%%%%%%%%%%%%%%%%%%%%%
\emph{(i) Model.---}We study a nano-electromechanical system called the \emph{single-electron shuttle}, which has 
been investigated theoretically~\cite{GorelikEtAlPRL1998, WeissZwergerEPL1999, BoeseSchoellerEPL2001, 
ArmourMacKinnonPRB2002, NordEtAlPRB2002, McCarthyPRB2003, NovotnyDonariniJauhoPRL2003, NovotnyEtAlPRL2004, 
UtamiEtAlPRB2006, NoceraEtAlPRB2011, PradaPlateroPRB2012}, experimentally~\cite{ParkEtAlNature2000, ErbeEtAlPRL2001, 
ScheibleBlickAPL2004, AyariEtAlNL2007, KimQinBlickAPL2007, MoskalenkoEtAlPRB2009, MoskalenkoEtAlNanoTech2009, 
KimQinBlickNJP2010, KimParkBlickPRL2010, KimPradaBlickACS2012, KoenigWeigAPL2012, WenEtAlNP2020} (for reviews see 
Refs.~\cite{ShekhterEtAlJPCM2003, GalperinRatnerNiztanJPCM2007, LaiZhangMaFP2015}) and recently also 
thermodynamically~\cite{TonekaboniLovettStaceArXiv2018, WaechtlerEtAlNJP2019, WaechtlerStrasbergSchallerPRAppl2019}. 
%The peculiar dynamics of the single-electron shuttle arises from the interplay of electrostatic and mechanical forces. 
Consider a quantum dot mounted on an oscillatory degree of freedom, which can move between two electron reservoirs 
(leads), see Fig.~\ref{fig sketch}. Proximity effects enhance (suppress) tunneling events of electrons 
whenever the dot is close (far) from the lead. Moreover, if electrons are on the dot, an electrostatic 
force acts in direction of the chemical potential bias (the voltage). Thus, the oscillator has 
the tendency to move with the bias whenever the dot is filled with electrons and, due to proximity effects, 
transport of electrons is enhanced due to the oscillation. Above a threshold voltage, this intrinsic feedback 
loop causes the oscillator to enter the regime of \emph{self-oscillations}~\cite{JenkinsPR2013}, even if it is 
damped by friction. This self-oscillation is responsible for the implementation of our thermodynamic cycle. 

We model the dynamics of the dot and oscillator semi-classically by a coupled Fokker-Planck and master 
equation~\cite{WaechtlerEtAlNJP2019}, which includes thermal fluctuations of the oscillator and allows us to 
study its entropy later on. In our regime of interest, quantum corrections to the oscillator are 
negligible~\footnote{For our numerical parameters, the thermal de Broglie wavelength 
$\lambda_\text{th} = h/\sqrt{2\pi mk_B T}$ roughly equals 50 femtometres, whereas shuttle oscillations happen in the 
few nanometres regime. }. In Appendix~\ref{sec derivation} we derive our equation of motion below starting from the 
quantum description~\cite{NovotnyDonariniJauhoPRL2003} and using phase space 
methods~\cite{FaddeevYakubovskiiBook2009, GardinerZollerBook2004}.

Let $P_q(x,v;t)$ be the probability density at time $t$ to find the oscillator at position $x$ (where $x=0$ 
defines the centre between the leads) with velocity $v$ and the dot with $q$ electrons. For simplicity we 
assume $q\in\{0,1\}$ (ultrastrong Coulomb blockade). This choice has little influence on the qualitative behaviour we 
are interested in, but we return to it at the end. Then, $P_q(x,v;t)$ obeys 
\begin{equation}\label{eq FPME}
 \frac{\partial P_q(x,v;t)}{\partial t} = L_q P_q(x,v;t) + \sum_{q'} R_{qq'}(x) P_{q'}(x,v;t),
\end{equation}
where we defined the following objects. First, 
\begin{equation}\label{eq generator FP}
 L_q \equiv -v\frac{\partial}{\partial x} + \frac{\partial}{\partial v}\left[\frac{k}{m}x + \frac{\gamma}{m}v - \frac{e\alpha V}{m}q + \frac{\gamma}{\beta m^2}\frac{\partial}{\partial v}\right]
\end{equation}
generates the oscillator movement as a function of the dot occupation $q$, where $k$ is the spring constant, $m$ 
the mass and the friction coefficient $\gamma = -F_\text{damp}/v$ results from a force $F_\text{damp}$ damping the 
oscillator in contact with an environment at inverse temperature $\beta = (k_BT)^{-1}$. The inverse distance 
$\alpha>0$ quantifies the strength of the electric field in between the leads and $V = (\mu_L-\mu_R)/e$ denotes the 
voltage ($e > 0$ is the elementary charge). The leads with chemical potential $\mu_L$ and $\mu_R$ are at the same 
temperature $T$. They influence the dynamics via the rate matrix $R(x) = R^L(x) + R^R(x)$, which can be split into 
contributions from the left and right lead and depends on the oscillator position $x$. Explicitly, the off-diagonal 
elements of $R^L(x)$ (the diagonal elements are fixed by probability conservation) describing the filling or 
depletion of the dot, respectively, read 
$R_{10}^L(x) = \Gamma_L(x) f_L[\epsilon(x)]$ and $R_{01}^L(x) = \Gamma_L(x) \{1-f_L[\epsilon(x)]\}$.
Here, $\Gamma_L(x) = \Gamma_0 e^{-x/\lambda}$ is an exponentially sensitive tunneling rate, $\Gamma_0$ a bare tunneling 
rate, $\lambda$ a characteristic tunneling distance and $f_L(\omega)=[e^{\beta(\omega-\mu_L)}+1]^{-1}$ the Fermi 
function. Importantly, the charging energy $\epsilon(x) = \epsilon_0 - e\alpha Vx$ of the filled dot is $x$-dependent 
($\epsilon_0$ is some effective on-site energy). Finally, the rate matrix $R^R(x)$ of the right lead is obtained 
from $R^L(x)$ by replacing $f_L$ by $f_R$ and by setting $\Gamma_R(x) = \Gamma_L(-x)$ (symmetric tunneling rates). 

We briefly discuss the thermodynamics of our autonomous model. The system (dot plus oscillator) is coupled to 
three baths: two electronic leads and the oscillator heat bath, labeled with a subscript `$O$' below. The heat 
flow up to time $t$ from bath $\nu\in\{L,R,O\}$ is denoted $Q_\nu(t)$. The first law reads 
\begin{equation}\label{eq 1st law auto}
 \Delta U_{DO}(t) = \sum_\nu Q_\nu(t) + W_\text{chem}(t),
\end{equation}
where $\Delta U_{DO} = U_{DO}(t) - U_{DO}(0)$ is the change in internal energy of the dot \emph{and} oscillator (we set 
the initial time to $t=0$) and $W_\text{chem}(t)$ is the chemical work associated to the transport of electrons 
(defined positive if electrons flow along the bias). Since all baths have the same temperature, the second law becomes 
\begin{equation}\label{eq 2nd law auto}
 \Delta S_{DO}(t) - \frac{1}{T} \sum_\nu Q_\nu(t) \ge 0
\end{equation}
with $S_{DO}(t)$ denoting the Gibbs-Shannon entropy of $P_q(x,v;t)$. 

We are only interested in \emph{average} thermodynamic quantities. Therefore, the above analysis is quite standard and 
detailed definitions are postponed to Appendix~\ref{sec thermo autonomous}. In our numerical simulations, however, we 
compute all quantities as averages over stochastic trajectories as detailed in Ref.~\cite{WaechtlerEtAlNJP2019}. 

\emph{(ii) Reduced dynamics.---}We now show how our autonomous model implements an idealized cycle in a certain 
parameter regime. Numerical simulations of Eq.~(\ref{eq FPME}) support our arguments. 

First, we want the oscillator to act like a \emph{work reservoir}, which is described by the ideal limit 
$m\rightarrow\infty$ while keeping $\omega = \sqrt{k/m}$ fixed~\cite{DeffnerJarzynskiPRX2013}. We argue below that 
this limit is actually `over-idealized,' but for now it is instructive to consider it. Then, the 
generator~(\ref{eq generator FP}) reduces to $L_q \rightarrow -v\partial_x + \partial_v\omega^2x$, which describes 
undisturbed motion of the oscillator according to the Hamiltonian $H_O(x,v) = mv^2/2 + kx^2/2$. If the initial 
condition is $P_q(x,v;0) = \C P_q(0)\delta(x-x_0)\delta(v)$, i.e., the oscillator starts at position $x_0$ with zero 
velocity, the state at time $t$ reads $P_q(x,v;t) = \C P_q(t)\delta(x-x_t)\delta(v-v_t)$ with $x_t = x_0\cos(\omega t)$ 
and $v_t = \dot x_t$. Thus, there is no backaction from the dot on the oscillator. However, the oscillator still 
influences the dot, which now obeys a \emph{time-dependent} master equation: 
\begin{equation}\label{eq ME}
 \frac{\partial\C P_q(t)}{\partial t} = \sum_{q'}R_{qq'}(x_t)\C P_{q'}(t).
\end{equation}
The solution of Eq.~(\ref{eq ME}), and quantities derived from it, is distinguished from the solution of the full dynamics~(\ref{eq FPME}) by using calligraphic symbols such as $\C P_q$. 

In reality, the above limit is too strong as it implies a constant oscillator energy: $dH_O(x_t,v_t)/dt = 0$. This is 
unphysical because Eq.~(\ref{eq ME}) predicts a finite energy flow into the oscillator (see below). Of course, 
in reality any oscillator mass is finite, albeit it can be very large. An adequate description is achieved by 
replacing $x_t$ with $\tilde x_t = a_t\cos(\omega t)$, where $a_t$ is the amplitude of the oscillator starting from 
$a_0 = x_0$. For large but finite mass $m$, $a_t$ varies \emph{slowly} in time, i.e., $a_t = x_0(1+\delta_t)$ with 
$|\dot \delta_t| \ll \omega$. Furthermore, for times $t$ such that $|\delta_t| \ll 1$, Eq.~(\ref{eq ME}) remains a 
good approximation while at the same time there is a \emph{finite change} in oscillator energy because the evaluation 
of $H_O(x_t,v_t) - H_O(x_0,v_0)$ involves terms like $\delta_t m x_0$, which can be large. 

The previous point is very important. The limit $m\rightarrow\infty$ is inconsistent, whereas the regime of 
finite but large $m$ makes our analysis consistent and non-trivial. From an analytical and numerical perspective, 
this is challenging as we can not rely on a steady state analysis of Eq.~(\ref{eq FPME}). To capture thermodynamic 
changes of the oscillator, we have to take into account its \emph{transient} dynamics. 

\begin{figure}
\includegraphics[width=0.48\textwidth,clip=true]{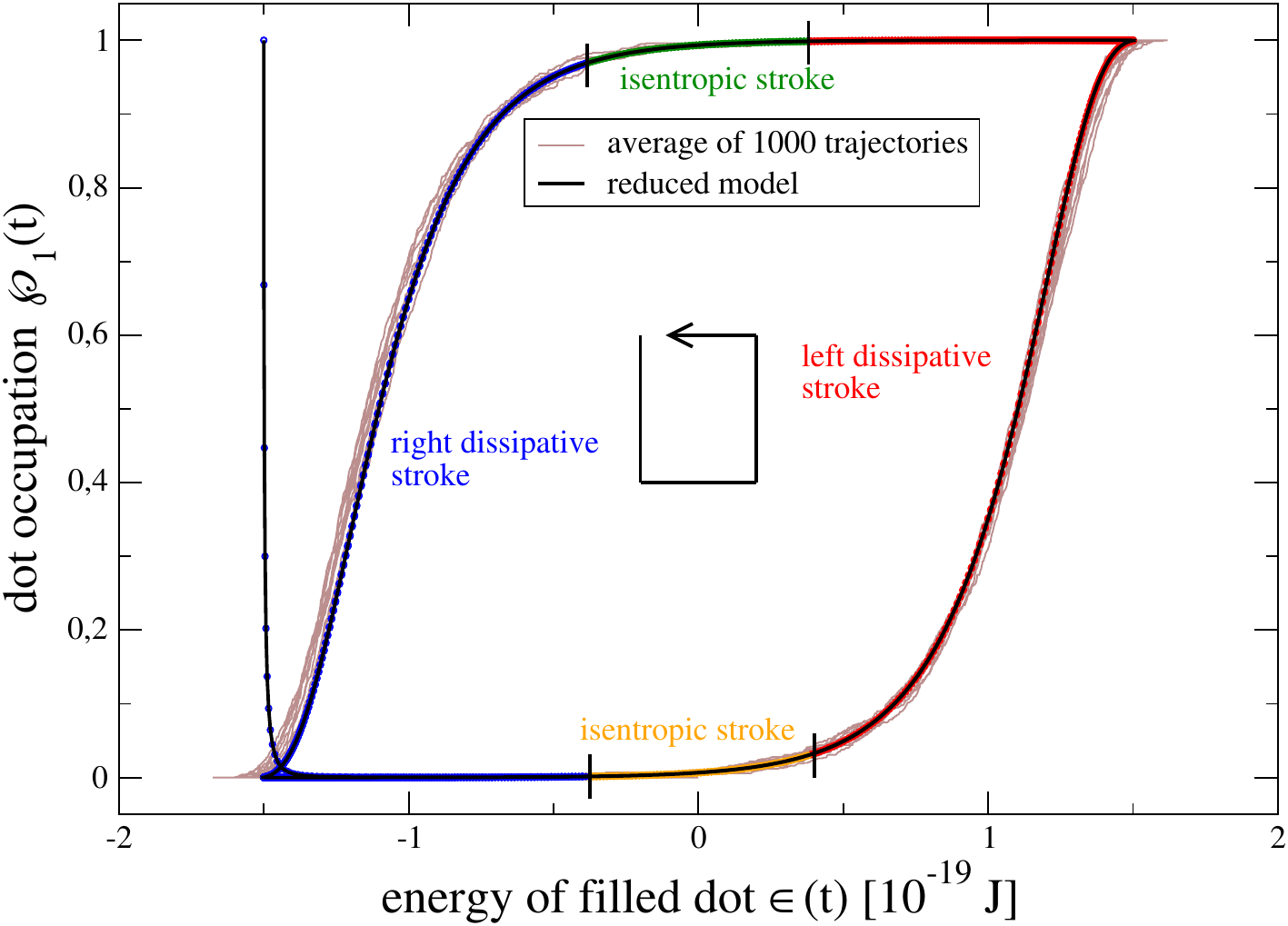}
\label{fig cyclelimit}
\caption{
Parametric plot of the dot occupation and energy $\epsilon(t) = \epsilon_0-\alpha eV x_t$, which is in 
one-to-one correspondence with the oscillator position $x_t$ ($\epsilon_0 = 0$ here). We compare the solution of 
Eq.~(\ref{eq ME}) with unperturbed oscillator trajectory $x_t = x_0 \cos(\omega t)$ (black and dotted) with the 
average of 1000 stochastic trajectories of the full dynamics~(\ref{eq FPME}) (thin brown) with identical initial 
condition for $t\in[0,250 {\rm ns}]$. The system quickly reaches a limit cycle, where the enclosed area measures the 
extracted work per cycle. }
\end{figure}

Finally, we justify the analysis in terms of a thermodynamic cycle divided into \emph{strokes} 
(cf.~Figs.~\ref{fig sketch} and~\ref{fig cyclelimit} and Appendix~\ref{sec partitioning}). First, the (approximately) 
periodic motion of the oscillator gives us the duration $\tau_\text{cycle} = 2\pi/\omega$ of one cycle. Next, if $x_0$ 
and $\lambda$ are chosen appropriately, the exponential sensitivity of the tunneling rates $\Gamma_{L/R}(x_t)$ 
justifies to neglect the influence of both leads when the dot is in the centre and to neglect the influence of the 
left (right) lead when the dot is on the right (left). The first case, determined by $\Gamma_{L/R}(x_t) \approx 0$, 
realizes an \emph{isentropic} stroke, where the dot does not change its state while its energy changes due to its 
movement in the potential bias. The second case, determined by $\Gamma_L(x_t) \gg 0$ \emph{or} $\Gamma_R(x_t) \gg 0$, 
realizes a \emph{dissipative} stroke, where, both, the state and energy of the dot changes. If parameters are 
fine-tuned such that the dot remains at temperature $T$, this stroke is isothermal. In general, however, the dot is 
out of equilibrium in our setup. 

Thus, we find that the cycle description is justified if
\begin{align}\label{eq conditions stroke}
 e^{|x_0|/\lambda} \gg 1, ~~~
 \tau_\text{isen}\Gamma_0 \exp\left[\frac{|x_0|}{\lambda}\sin\left(\frac{\omega\tau_\text{isen}}{2}\right)\right] \ll 1.
\end{align}
The first condition in Eq.~(\ref{eq conditions stroke}) is necessary to neglect the effect of the opposite, remote lead 
during the dissipative strokes. The second condition involves the duration $\tau_\text{isen}$ of the isentropic 
strokes, which depends on other parameters of the model. It is derived in detail in Appendix~\ref{sec partitioning}. 

We are particularly interested in the properties of our device as a function of the oscillator mass $m$ (keeping 
$\omega$ fixed) and the friction coefficient $\gamma$. The other parameters are based on reasonable estimates from 
Refs.~\cite{KimQinBlickAPL2007, KimQinBlickNJP2010, KimParkBlickPRL2010, KimPradaBlickACS2012, PradaPlateroPRB2012}, 
precisely listed in Appendix~\ref{sec parameter choice}. For them we find a duration 
$\tau_\text{isen} = \tau_\text{cycle}/12$ of the isentropic stroke in unison with 
condition~(\ref{eq conditions stroke}) with a total cycle time $\tau_\text{cycle}\approx 25$~ns. 

\emph{(iii) Reduced thermodynamics.---}We start with the analysis of Eq.~(\ref{eq ME}), distinguished by 
calligraphic symbols. For now, we ignore the fact that Eq.~(\ref{eq ME}) follows from an underlying autonomous 
model---instead, we assume that the time-dependent rate matrix is generated by an ideal work reservoir as 
conventionally done in thermodynamic cycle analyses~\cite{VinjanampathyAndersCP2016, KosloffRezekEntropy2017, 
GhoshEtAlBook2018, FeldmannPalaoBook2018, LevyGelbwaserKlimovskyBook2018, DeffnerCampbellBook2019}. 
Then, mechanical work becomes 
\begin{align}\label{eq work per cycle}
 \C W_{\rm mech}(t) = \int_0^t dt' \C P_1(t') \frac{\partial\epsilon(x_{t'})}{\partial t'}.
\end{align}
For a single cycle the work equals the area enclosed by the limit cycle trajectory (counted positive 
in clockwise direction in Fig.~\ref{fig cyclelimit}), similar to a $p$--$V$ diagram in traditional cycles. 

As before, there are heat flows $\C Q_\nu(t)$ from lead $\nu$ and chemical work $\C W_\text{chem}(t)$ such that 
the first law reads 
\begin{equation}\label{eq 1st law red}
 \Delta\C U_D(t) = \C Q_L(t) + \C Q_R(t) + \C W_{\rm chem}(t) + \C W_{\rm mech}(t).
\end{equation}
Here, $\C U_D(t) = \epsilon(x_t) \C P_1(t)$ is the internal energy of the dot. Furthermore, denoting by $\C S_D(t)$ 
the Gibbs-Shannon entropy of %the dot distribution 
$\C P_q(t)$, the second law reads 
\begin{equation}\label{eq 2nd law red}
 \Delta\C S_D(t) - \frac{1}{T}[\C Q_L(t) + \C Q_R(t)] \ge 0.
\end{equation}
This analysis follows again from standard considerations and explicit expressions are thus only displayed in 
Appendix~\ref{sec thermo reduced}. Note that Eqs.~(\ref{eq 1st law red}) and~(\ref{eq 2nd law red}), while 
mathematically true, need not coincide with the thermodynamics of the autonomous model. In general, the dot dynamics 
predicted by both methods differ, i.e., $\C P_q(t) \neq P_q(t) \equiv \int dx dv P_q(x,v;t)$. 

We now return to the first and second law of the autonomous model, Eqs.~(\ref{eq 1st law auto}) 
and~(\ref{eq 2nd law auto}), rewritten as (dropping the $t$-dependence for simplicity) 
\begin{align}
 & \Delta U_D = Q_L + Q_R + W_\text{chem} + Q_O - \Delta U_O, \label{eq 1st law auto 2} \\
 & \Delta S_D - \frac{Q_L}{T} - \frac{Q_R}{T} + \Delta S_{O|D} - \frac{Q_O}{T} \ge 0. \label{eq 2nd law auto 2}
\end{align}
Here, the oscillator energy $U_O$ equals the expectation value of its Hamiltonian $H_O(x,v)$ and 
$S_{O|D} = S_{DO} - S_D$ is the conditional entropy. The necessary conditions for the thermodynamic 
laws~(\ref{eq 1st law red}) and~(\ref{eq 2nd law red}) of the reduced model to coincide with 
Eqs.~(\ref{eq 1st law auto 2}) and~(\ref{eq 2nd law auto 2}) follow as 
\begin{equation}
 \frac{Q_O}{T} = 0 ~~~ \text{and} ~~~ \Delta S_{O|D} = 0.
\end{equation}
Of course, on top of that, we also need $\C P_q(t) = P_q(t)$. 
%, which implies $Q_\nu = \C Q_\nu$ ($\nu\in\{L,R\}$), etc. 

\begin{figure}
\includegraphics[width=0.48\textwidth,clip=true]{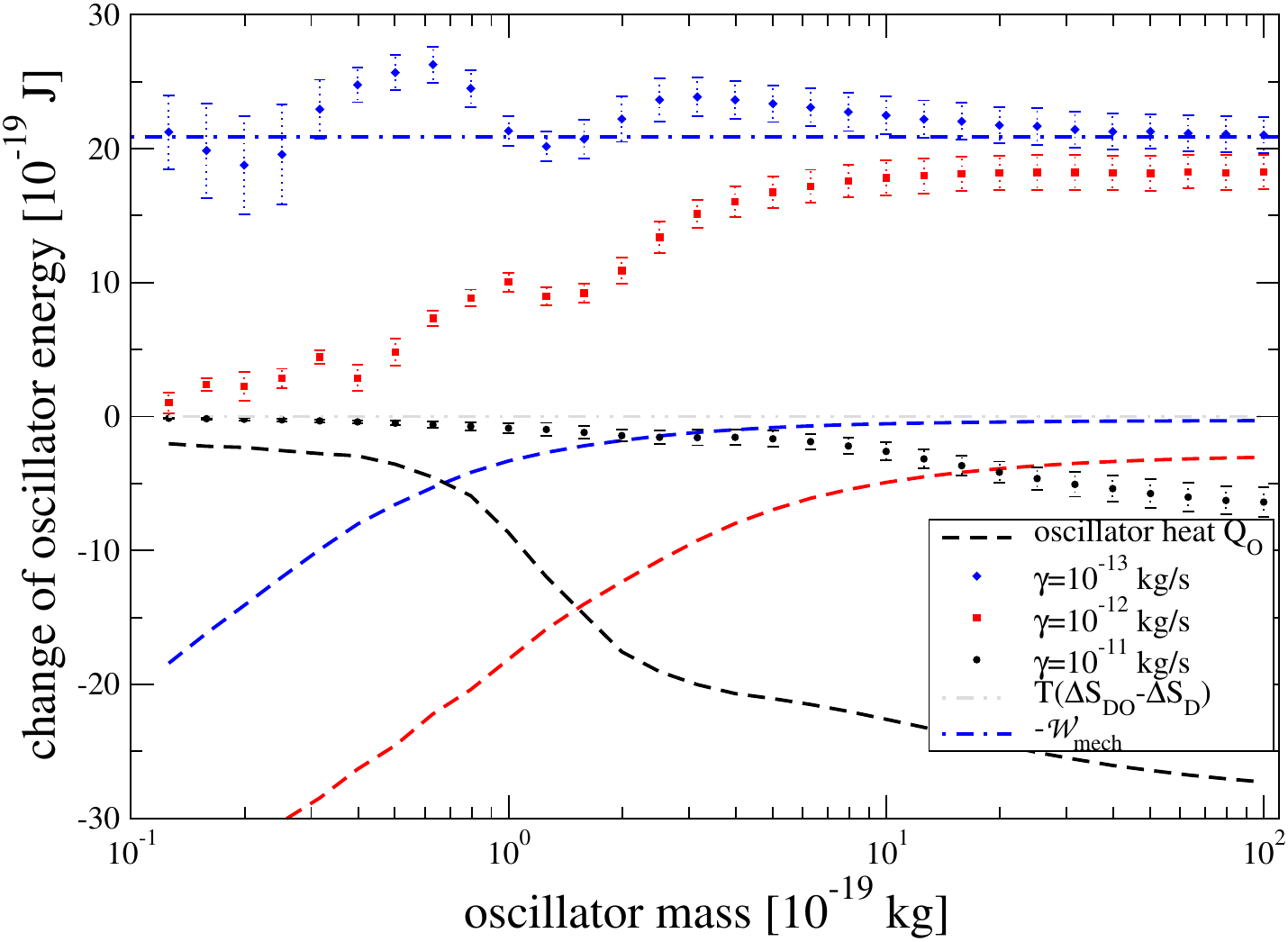}
\label{FIG:heatandwork}
\caption{
Average energy change of the oscillator $\Delta U_O$ (with error bars marking the statistical error) over the time 
interval $[0,250 {\rm ns}]$ versus mass for different friction coefficients (in logarithmic scale). Dashed curves of 
matching color correspond to the heat flow $Q_O$ from the oscillator bath. As $\gamma\to 0$ and $m\to\infty$, 
$\Delta U_O$ matches $-\C W_\text{mech}$ (dash-dotted blue line) and $Q_O$ and $\Delta S_{O|D}$ become negligible. }
\end{figure}

Even if all parameters are kept finite in the original model, the above conditions can be satisfied to good 
approximation. First, for large mass $m$, keeping $k/m$ fixed, the dynamics is well-described by 
Eq.~(\ref{eq ME}), i.e., $\C P_q(t) \approx P_q(t)$. We checked this numerically for multiple parameters, see 
Fig.~\ref{fig cyclelimit} for a particular example. Furthermore, the solution of Eq.~(\ref{eq FPME}) remains 
approximately of the form $P_q(x,v;t) \approx \C P_q(t)\delta(x-x_t)\delta(v-v_t)$, i.e., the oscillator state has 
low entropy for long times, which implies $\Delta S_{O|D} \approx 0$
%Hence, $S_{DO}(t) - S_D(t) \approx 0$ and therefore $S_{O|D}(t) \approx 0$ (see the 
(dash-dotted grey line in Fig.~\ref{FIG:heatandwork}). 

The previous argument is not yet sufficient to conclude that $Q_O/T \approx 0$. Instead, the heat flow $Q_O$ 
is controlled by the friction coefficient $\gamma$. Thus, on top of the large $m$ regime, we also require small 
$\gamma$. Then, Eqs.~(\ref{eq 1st law auto 2}) and~(\ref{eq 2nd law auto 2}) coincide with Eqs.~(\ref{eq 1st law red}) 
and~(\ref{eq 2nd law red}). In this limit, the oscillator resembles a perfect work reservoir or \emph{battery}. 

These arguments are exemplified in Fig.~\ref{FIG:heatandwork}. First, for increasing $m$, we see that we obtain 
$\Delta U_O - Q_O \approx -\C W_\text{mech}$. Second, we observe for decreasing $\gamma$ that $Q_O\rightarrow0$. Hence, 
$\Delta U_O = -\C W_\text{mech}$ and the mechanical work computed from the reduced dynamics~(\ref{eq ME}) is 
stored as \emph{extractable} energy in the oscillator since $T|\Delta S_O| \ll |\Delta U_O|$. 

The question remains whether also the cycle analysis matches this picture. However, 
if condition~(\ref{eq conditions stroke}) is satisfied, then the dynamics of the master equation~(\ref{eq ME}) matches 
the cycle dynamics. This is constructed by time-evolving the state $\C P_q(t)$ only with respect to the rate matrix 
$R^L(x_t)$ [$R^R(x_t)$] during the left [right] dissipative stroke and by using the identity map for the 
isentropic strokes. In this regime, the thermodynamic quantities in Eqs.~(\ref{eq 1st law red}) 
and~(\ref{eq 2nd law red}) coincide \emph{by construction} with the cycle analysis, 
demonstrated in detail in Appendix~\ref{sec cyle analysis reduced}. 

\emph{Experimental feasibility.---}Can thermodynamic cycles realistically be used to analyse nanoscale engines? 
Based on our parameter choice (see Appendix~\ref{sec parameter choice}), we find the following. First, the 
experimentally used mass $m$ is already sufficiently large, albeit an increase in it would still be beneficial. 
Second, to mimic an ideal work reservoir, the friction $\gamma$ needs to be about one order of magnitude smaller than 
typical experimental values. This could be in reach with current technologies. Third, as stated in the SM, our 
numerics is based on a large voltage $V$. This likely invalidates the ultrastrong Coulomb blockade assumption and 
raises the final question: is a cycle with a \emph{single-electron} working fluid realistic? 

To answer it, we estimate the number $N$ of electrons contributing to the transport in the `bias window' 
$\Delta E \equiv eV$. The electrostatic energy of the dot is $E_D = (eN)^2/(2C)$, where $C \approx 4\pi \epsilon_0 d$ 
is the self capacitance of the dot mounted on the nanopillar with diameter $d$~\footnote{This is a simplified 
estimate, which we believe to be sufficient for our purposes~\cite{NazarovBlanterBook2009}. Determining the 
exact electrostatic energy of a quantum dot is quite complicated, see, e.g., Ref.~\cite{RanjanEtAlPRB2002}}. Equating 
$\Delta E = E_D$, we obtain $N^2 \approx 3.5\cdot 10^{10}\cdot d/$m for our parameter choice. 
For $N = 1$ (the case considered here), this requires a nanopillar with a diameter of 30~pm. This is smaller than 
the radius of a hydrogen atom and impossible to fabricate. Hence, we assume a diameter of 5~nm for stability 
reasons, which is optimistic compared to experimental values of $d = 60$~nm~\cite{KimQinBlickNJP2010}\footnote{For a 
60~nm pillar, our estimate predicts $N\approx 45$ electrons. Given the simplicity of our estimate, this matches well 
the reported 100 electrons shuttled per cycle in Ref.~\cite{KimQinBlickNJP2010} if one takes into account that this 
experiment was done at room temperature whereas we assumed $T =1$~K. }. Then, we obtain the estimate $N\approx 13$. 
We conclude that an autonomous implementation of a thermodynamic cycle with a single-electron working medium seems 
experimentally \emph{impossible}, but $N\gtrsim 10$ electrons could suffice. 

Our conclusions seem to remain for other experimental platforms~\footnote{An early 
experiment~\cite{AyariEtAlNL2007} with nanowires of $20-400$~nm diameter reports self-oscillations for 
voltages around $V\gtrsim 100$~V with $N\approx50.000$ electrons (indirectly inferred by assuming a shuttling period 
of $15~\mu$s). In Refs.~\cite{MoskalenkoEtAlPRB2009, MoskalenkoEtAlNanoTech2009} for a 20~nm gold-nanoparticle-shuttle, 
shuttling above a voltage of 3~V with $\approx 20$ electrons was reported, which matches well our estimate. Finally, 
Ref.~\cite{KoenigWeigAPL2012} uses a gold island with size $35\times270\times40$~nm to shuttle $N\approx200$ electrons 
per cycle above a threshold voltage $V\gtrsim 5$~V.}, 
but a recent experiment~\cite{WenEtAlNP2020} reports sustained oscillations of a suspended carbon nanotube for $N=1$ 
electron. However, in view of our Fig.~\ref{fig sketch}, the nanotube oscillates \emph{vertically} (i.e., 
up--down instead of left--right), which makes the identification of \emph{strokes} unclear. Nevertheless, 
it remains an intriguing question whether the nanotube acts like an ideal work reservoir. 

We remark that we have not shown how to implement a Carnot or Otto cycle. Our cycle is driven by a 
voltage instead of temperature bias, converting chemical work into mechanical work. 
%This is nevertheless of technological relevance: 
%a car engine does nothing but converting chemical work into mechanical work in analogy to our model. 
Autonomously realizing Carnot and Otto cycles faces additional challenges and remains open. 
%Since the Otto cycle assumes a strict separation 
%between strokes exchanging \emph{only} work \emph{or} heat, this faces additional challenges. 

\emph{Conclusions.---}We demonstrated the potential of self-oscillating engines to address problems of foundational 
and practical relevance. 
%We believe that these models pave the way for fruitful future 
%research avenues, in unison with other recent works on devices with self-oscillating or rotational 
%components~\cite{TonekaboniLovettStaceArXiv2018, WaechtlerEtAlNJP2019, WaechtlerStrasbergSchallerPRAppl2019, 
%FilligerReimannPRL2007, WangVukovicKralPRL2008, AlickiGelbwaserKlimovskySzczygielskiJPA2015, SerraGarciaEtAlPRL2016, 
%AlickiJPA2016, ChiangEtAlPRE2017, RouletEtAlPRE2017, AlickiGelbwaserKlimovskyJenkinsAP2017, 
%SeahNimmrichterScaraniNJP2018, AlickiEtAlArXiv2020}. 
These models could pave the way for fruitful future research avenues, as evidenced also by other recent 
studies~\cite{TonekaboniLovettStaceArXiv2018, WaechtlerEtAlNJP2019, WaechtlerStrasbergSchallerPRAppl2019, 
FilligerReimannPRL2007, WangVukovicKralPRL2008, AlickiGelbwaserKlimovskySzczygielskiJPA2015, SerraGarciaEtAlPRL2016, 
AlickiJPA2016, ChiangEtAlPRE2017, RouletEtAlPRE2017, AlickiGelbwaserKlimovskyJenkinsAP2017, 
SeahNimmrichterScaraniNJP2018, AlickiEtAlArXiv2020}. 

\emph{Acknowledgements.---}PS is financially supported by the DFG (project STR 1505/2-1), the Spanish Agencia Estatal 
de Investigaci\'on, project PID2019-107609GB-I00, the Spanish MINECO FIS2016-80681-P (AEI/FEDER, UE), and Generalitat 
de Catalunya CIRIT 2017-SGR-1127. CWW and GS acknowledge support by the DFG through Project No. BR1528/8-2. 
CWW acknowledges support from the Max-Planck Gesellschaft via the MPI-PKS Next Step fellowship.

%%%%%%%%%%%%%%%%%%%%%%%%%%%%%%%%%%%%%%%%%%%%%%%%%%%%%%%%%%%%%%%%%%%%%%%%%%%%%%%%%%%%%%%%%%%%%%%%%%%%%%%%%%%%%%%%%%%%%%%%

\bibliography{/home/philipp/Documents/references/books,/home/philipp/Documents/references/open_systems,/home/philipp/Documents/references/thermo,/home/philipp/Documents/references/info_thermo,/home/philipp/Documents/references/general_QM,/home/philipp/Documents/references/math_phys,/home/philipp/Documents/references/equilibration}
%\bibliography{/home/wiwi/Documents/references/books,/home/wiwi/Documents/references/open_systems,/home/wiwi/Documents/references/thermo,/home/wiwi/Documents/references/info_thermo,/home/wiwi/Documents/references/general_QM,/home/wiwi/Documents/references/math_phys,/home/wiwi/Documents/references/general_refs,/home/wiwi/Documents/references/equilibration}
%\bibliography{books,open_systems,thermo,general_QM,info_thermo,math_phys}

\onecolumngrid
%%%%%%%%%%%%%%%%%%%%%%%%%%%%%%%%%%%%%%%%%%%%%%%%%%%%%%%%%%%%%%%%%%%%%%%%%%%%%%%%%%%%%%%%%%%%%%%%%%%%%%%%%%%%%%%%%%%%%%%%
\appendix
%%%%%%%%%%%%%%%%%%%%%%%%%%%%%%%%%%%%%%%%%%%%%%%%%%%%%%%%%%%%%%%%%%%%%%%%%%%%%%%%%%%%%%%%%%%%%%%%%%%%%%%%%%%%%%%%%%%%%%%%
\section{Derivation of the coupled Fokker-Planck and master equation}
\label{sec derivation}

Our starting point is the full quantum master equation for the combined dot-oscillator system as derived in 
Ref.~\cite{NovotnyDonariniJauhoPRL2003}: 
\begin{align}
 \frac{\partial}{\partial t}\hat\rho(t) 
 =&~ (\C L_\text{coh} + \C L_\text{drive} + \C L_\text{damp}) \hat\rho(t), \label{eq QME} \\
 \C L_\text{coh}\hat\rho =&~ \frac{1}{\rm i\hbar}
 [H_\text{osc} + \epsilon_0 \hat c_0^\dagger \hat c_0 - e E \hat x\hat c_0^\dagger \hat c_0,\hat\rho], \\
 \C L_\text{drive}\hat\rho =& -\frac{\Gamma_L}{2}\left(\hat c_0\hat c_0^\dagger e^{-2\hat x/\lambda'} \hat\rho 
 - 2\hat c_0^\dagger e^{-\hat x/\lambda'}\hat\rho e^{-\hat x/\lambda'}\hat c_0 
 + \hat\rho e^{-2\hat x/\lambda'}\hat c_0\hat c_0^\dagger\right) \\
 &- \frac{\Gamma_R}{2}\left(\hat c_0^\dagger\hat c_0 e^{2\hat x/\lambda'} \hat\rho 
 - 2\hat c_0 e^{\hat x/\lambda'}\hat\rho e^{\hat x/\lambda'}\hat c^\dagger_0 
 + \hat\rho e^{2\hat x/\lambda'}\hat c_0^\dagger\hat c_0\right), \nonumber \\
 \C L_\text{damp}\hat\rho =& -\frac{\rm i\gamma'}{2\hbar}[\hat x,\{\hat p,\hat\rho\}] 
 - \frac{\gamma' m\omega}{\hbar}\left(\bar N + \frac{1}{2}\right) [\hat x,[\hat x,\hat\rho]].
\end{align}
Here, we mostly followed the notation of Ref.~\cite{NovotnyDonariniJauhoPRL2003}, but explicitly denoted operators with 
a hat for convenience. The notation of the main text is obtained after identifying $H_\text{osc} = H_O$, 
$\lambda' = 2\lambda$, $E = \alpha V$ and $\gamma' = \gamma/m$ (note that the friction coefficient $\gamma$ in the main 
text does not have the dimension of a rate in contrast to $\gamma'$). Furthermore, $\hat c_0^\dagger$ ($\hat c_0$) 
creates (annihilates) an electron on the dot and $\bar N = (e^{\beta\hbar\omega}-1)^{-1}$ is the Bose-Einstein 
distribution. Finally, we point out that the master equation in Ref.~\cite{NovotnyDonariniJauhoPRL2003} was derived in 
the `high-bias' limit, which allowed them to replace the Fermi functions by $f_L(\omega) \approx 1$ and 
$f_R(\omega) \approx 0$ for all dot energies $\omega$. For simplicity in the presentation and in unison with 
Ref.~\cite{NovotnyDonariniJauhoPRL2003} we keep the Fermi function out of the discussion here. 

For the reasons spelled out in Footnote~[62] of the main text, we are interested in the classical limit of the 
quantum master equation above. This is most conveniently derived by considering the time-evolution of the Wigner 
function $W(x,p)$ of the oscillator (where $p$ is its momentum) and by taking the formal limit 
$\hbar\rightarrow0$~\cite{FaddeevYakubovskiiBook2009}. Moreover, we are only interested in the occupation probabilities 
of the dot as coherences between the empty and filled state of the dot are prohibited since they correspond to 
superpositions of different charged states. Thus, we define 
\begin{equation}
 W_q(x,p) = \frac{1}{\hbar\pi} \int_{-\infty}^\infty dy \langle x-y,q|\hat\rho|x+y,q\rangle e^{2\rm ipy/\hbar},
\end{equation}
where $q\in\{0,1\}$ denotes the number of electrons on the dot. The time-evolution of $W_q(x,p)$ can then be derived 
from Eq.~(\ref{eq QME}) by using operator correspondence rules, which can be readily checked for consistency in 
textbooks~\cite{GardinerZollerBook2004}. Examples are 
\begin{equation}\label{eq correspondence rules}
 \hat x\hat\rho \mapsto \left(x + \frac{\rm i\hbar}{2}\frac{\partial}{\partial p}\right) W_q(x,p), ~~~ 
 \hat p\hat\rho \mapsto \left(p - \frac{\rm i\hbar}{2}\frac{\partial}{\partial x}\right) W_q(x,p), ~~~ \dots, 
\end{equation}
from which we can already confirm the position-momentum commutation relation. Multiplication from the right with these operators follows from Hermitian conjugation, as does a combination of them (provided one minds the correct ordering). 

We find for the first term in Eq.~(\ref{eq QME}) that 
\begin{equation}\label{eq L1}
 \C L_\text{coh}\hat\rho \mapsto 
 \left[-\frac{\partial}{\partial x} \frac{p}{m} + \frac{\partial}{\partial p} (kx - eEq)\right]W_q(x,p),
\end{equation}
without any need to take the limit $\hbar\rightarrow0$. After setting $v = p/m$, Eq.~(\ref{eq L1}) reproduces the 
first, second and fourth term of the generator $L_q$ defined in Eq.~(2) of the main text. 

Next, we consider the second term $\C L_\text{drive}\hat\rho(t)$ in Eq.~(\ref{eq QME}). The exponential factors 
$e^{\hat x/\lambda'}$ in it make the mapping complicated in principle, but remember that we are only interested in 
the limit $\hbar\rightarrow0$. Since $\hbar$ appears nowhere explicitly in $\C L_\text{drive}$, we can directly use 
$\lim_{\hbar\rightarrow0} \hat x\hat\rho \mapsto x W_q(x,p)$, which follows from Eq.~(\ref{eq correspondence rules}). 
Thus, we can set, e.g., $\lim_{\hbar\rightarrow0} e^{\hat x/\lambda'}\hat\rho \mapsto e^{x/\lambda'} W_q(x,p)$ 
and we obtain 
\begin{equation}
 \begin{split}
  \lim_{\hbar\rightarrow0}\C L_\text{drive}\hat\rho \mapsto& -\frac{\Gamma_L}{2}e^{-2x/\lambda'} 
  \left(\hat c_0\hat c_0^\dagger W(x,p) - 2\hat c_0^\dagger W(x,p) \hat c_0 + W(x,p)\hat c_0\hat c_0^\dagger\right) \\
  &- \frac{\Gamma_R}{2}e^{2x/\lambda'}
  \left(\hat c_0^\dagger\hat c_0 W(x,p) - 2\hat c_0 W(x,p)\hat c^\dagger_0 + W(x,p)\hat c_0^\dagger\hat c_0\right). \\
 \end{split} 
\end{equation}
Here, we have not yet taken matrix element $|q\rangle$ in the dot basis. Doing so reveals that 
\begin{equation}
 \lim_{\hbar\rightarrow0}\C L_\text{drive}\hat\rho \mapsto \sum_{q'} R_{qq'}(x) W_{q'}(x,p),
\end{equation}
where $R_{qq'}(x)$ is the rate matrix defined in the main text in the high bias limit (as discussed above). 

Finally, the last term $\C L_\text{damp} \hat\rho(t)$ in Eq.~(\ref{eq QME}) simply describes the dynamics of a 
damped harmonic oscillator. It reduces to the third and fifth term of the generator $L_q$ defined in Eq.~(2) of the 
main text for $\hbar\rightarrow0$, after paying attention to the fact that $\gamma' = \gamma/m$. Thus, after setting 
$P_q(x,v;t) \equiv W_q(x,mv)$, we obtain Eq.~(1) of the main text as the classical limit of the quantum master 
equation derived in Ref.~\cite{NovotnyDonariniJauhoPRL2003}. Note that in the classical limit $P_q(x,v;t) = W_q(x,mv)$ 
has no negativities and becomes a well defined probability density. 

%%%%%%%%%%%%%%%%%%%%%%%%%%%%%%%%%%%%%%%%%%%%%%%%%%%%%%%%%%%%%%%%%%%%%%%%%%%%%%%%%%%%%%%%%%%%%%%%%%%%%%%%%%%%%%%%%%%%%%%%
\section{Precise thermodynamic definitions for the autonomous model}
\label{sec thermo autonomous}

We define the internal energy and entropy of the combined dot-oscillator system as 
\begin{equation}
 U_{DO}(t) \equiv \sum_q\int dxdv \left[\frac{mv^2}{2} + \frac{kx^2}{2} + \epsilon(x)q\right] P_q(x,v;t), ~~~
 S_{DO}(t) \equiv -k_B\sum_q\int dxdv P_q(x,v;t)\ln\left[\frac{\hbar}{m}P_q(x,v;t)\right].
\end{equation}
We remark that the factor $\hbar/m$, which ensures that the argument of the logarithm is dimensionless, cancels out 
whenever we take differences of $S_{DO}(t)$. This is always the case in the following. 
Furthermore, the instantaneous heat flow from lead $\nu\in\{L,R\}$ is composed out of an energy and a particle current: 
$\dot Q_\nu(t) = J_\nu(t) - \mu_\nu I_\nu(t)$. They are defined as 
\begin{equation}
 J_\nu(t) \equiv \sum_{q,q'} \int dxdv (\epsilon_0-\alpha eVx)q R^\nu_{q,q'}(x) p_{q'}(x,v;t), ~~~
 I_\nu(t) \equiv \sum_{q,q'} \int dxdv q R_{q,q'}^\nu(x) p_{q'}(x,v;t).
\end{equation}
The instantaneous heat flow from the oscillator bath $\dot Q_O(t)$ only has an energy component: 
\begin{equation}
 \dot Q_O(t) = \sum_q \int dxdv \left(\frac{mv^2}{2} + \frac{kx^2}{2} - \alpha eVx q\right) L_q P_q(x,v;t). 
\end{equation}
The heat flows appearing in the first law~(3) in the main text follow by integration: 
$Q_\nu(t) = \int_0^t dt'\dot Q_\nu(t')$, $\nu\in\{L,R,O\}$. Finally, the chemical work is defined as 
\begin{equation}
 W_\text{chem}(t) \equiv \int_0^t dt' \left[\mu_L I_L(t') + \mu_R I_R(t')\right].
\end{equation}

%%%%%%%%%%%%%%%%%%%%%%%%%%%%%%%%%%%%%%%%%%%%%%%%%%%%%%%%%%%%%%%%%%%%%%%%%%%%%%%%%%%%%%%%%%%%%%%%%%%%%%%%%%%%%%%%%%%%%%%%
\section{Partitioning the cycle into strokes}
\label{sec partitioning}

\begin{figure}
\includegraphics[width=0.50\textwidth,clip=true]{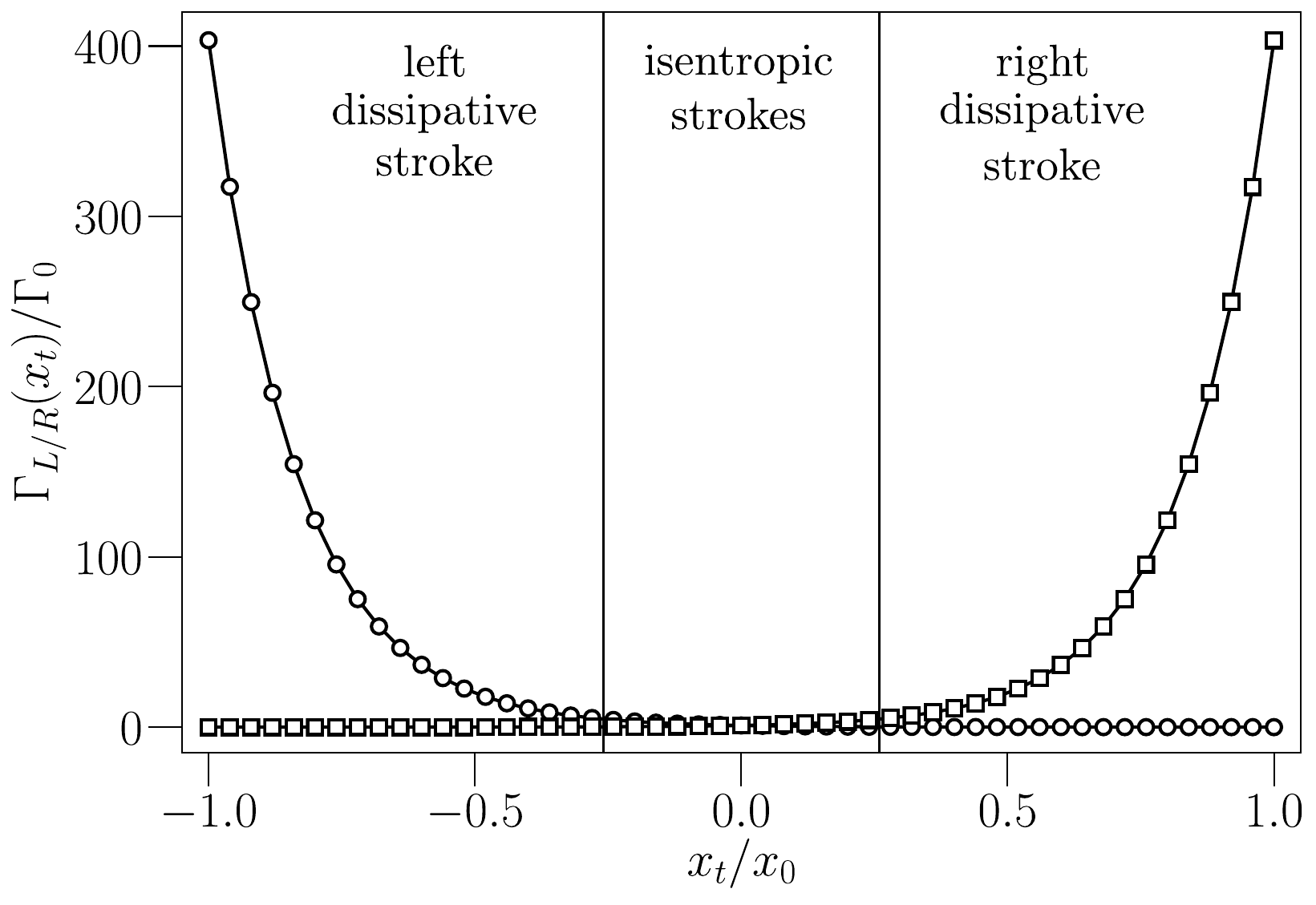}
\label{fig strokes gamma}
\caption{Plot of the bare tunneling rates $\Gamma_L(x_t)$ (circles) and $\Gamma_R(x_t)$ (squares) as a function 
of the oscillator trajectory $x_t = x_0\cos(\omega t)$ obtained in the limit of an ideal work reservoir for the 
numerical parameters specified in Sec.~\ref{sec parameter choice}. The regimes where we identify the different strokes 
are separated by vertical lines. We remark that the precise numerical definition of the strokes requires to introduce a 
threshold value and is therefore only fixed with respect to that value. To be on the safe side, we choose our 
threshold value such that the integral~(\ref{eq integral}) is much smaller than one. Then, at the boundary of the 
isentropic stroke, the dot is still hardly coupled to any of the leads, and small changes of the cycle durations will 
have no net effect.}
\end{figure}

The emergence of different strokes in our analysis arises from the sensitivity of the bare tunneling rates 
$\Gamma_{L/R}(x_t)$ with respect to the changing position $x_t$ of the oscillator as sketched in 
Fig.~\ref{fig strokes gamma}. If $\Gamma_L(x_t) \gg \Gamma_R(x_t) \approx 0$ 
[$\Gamma_R(x_t) \gg \Gamma_L(x_t) \approx 0$], the oscillator is on the left [right] and we can neglect the influence 
of the opposite lead, which defines the respective dissipative strokes. If 
$\Gamma_L(x_t)\approx\Gamma_R(x_t) \approx 0$, the oscillator is in the middle, which defines the isentropic strokes. 
Whether these strokes can be identified and how long they last depends on the precise choice of numerical 
parameters. Below, we estimate the time $\tau_\text{isen}$ that the oscillator spends in the centre with negligible 
influence from both leads, i.e., we ask when is $\Gamma_L(x_t)\approx\Gamma_R(x_t) \approx 0$. This gives us the 
second condition in Eq.~(6) in the main text.

The time evolution of the dot is given by the time-ordered exponential of the rate matrix 
appearing in Eq.~(5) in the main text and we seek to find the time-intervals in which this is approximately equal to 
the identity, i.e., the state of the dot remains unchanged. The length of this `isentropic' time interval is denoted 
$\tau_\text{isen}$ in the following. Clearly, by symmetry these time-interval are centered around the times 
$\pi(n+\frac{1}{2})/\omega$ ($n\in\mathbb{N}$) when the oscillator is in the centre around $x_t\approx0$ [note that 
the initial condition for the oscillator is $(x_0,v_0) = (|x_0|,0)$, see Sec.~\ref{sec parameter choice}]. 
To determine the length $\tau$ of, say, the first time-interval, we demand that 
\begin{equation}\label{eq integral}
 I \equiv \int\limits_{\frac{\pi}{2\omega}-\frac{\tau_\text{isen}}{2}}^{\frac{\pi}{2\omega}+\frac{\tau_\text{isen}}{2}} dt
 \left[\Gamma_L(x_t) + \Gamma_R(x_t)\right] \ll 1.
\end{equation}
Here, $\Gamma_L(x_t) = \Gamma_0 e^{-x_t/\lambda} = \Gamma_R(-x_t)$ are the bare tunneling rates for the left and right 
lead \emph{ignoring} the influence of the Fermi functions. Since the Fermi functions are always smaller than one, 
neglecting them only underestimates the length $\tau_\text{isen}$. 

Nevertheless, an exact analytical evaluation of the integral~(\ref{eq integral}) is still not possible. Therefore, we 
make further approximations. First, we assume the necessary requirement $\exp(|x_0|/\lambda) \gg 1$, i.e., the first 
condition of Eq.~(6) in the main text, to be satisfied. Together with our choice for the initial state of the 
oscillator, we simplify 
\begin{equation}
 I \approx \int\limits_{\frac{\pi}{2\omega}-\frac{\tau_\text{isen}}{2}}^{\frac{\pi}{2\omega}} \Gamma_L(x_t)dt + 
 \int\limits_{\frac{\pi}{2\omega}}^{\frac{\pi}{2\omega}+\frac{\tau_\text{isen}}{2}} \Gamma_R(x_t)dt \ll 1.
\end{equation}
This step undestimates the value of $I$, but this is well compensated by the next crude approximation, where we replace 
$\Gamma_L(x_t)$ and $\Gamma_R(x_t)$ by their maximum values taken at the boundaries for 
$t = \frac{\pi}{2\omega}-\frac{\tau_\text{isen}}{2}$ and $t = \frac{\pi}{2\omega}+\frac{\tau_\text{isen}}{2}$, 
respectively. Then, 
\begin{equation}
 I \approx \frac{\tau_\text{isen}}{2}\Gamma_L\left(x_{\pi/2\omega - \tau_\text{isen}/2}\right) 
 + \frac{\tau_\text{isen}}{2}\Gamma_R\left(x_{\pi/2\omega + \tau_\text{isen}/2}\right)
 = \tau_\text{isen}\Gamma_0\exp\left[\frac{|x_0|}{\lambda}\sin\left(\frac{\omega\tau_\text{isen}}{2}\right)\right],
\end{equation}
where we used the identity $\cos(\pi/2 \pm x) = \mp \sin(x)$. Now, the requirement that 
$I\ll1$ gives the second condition in Eq.~(6) in the main text. 

For the numerical parameters listed in Sec.~\ref{sec parameter choice} below, we find a partition into strokes as 
summarized in Table~\ref{table strokes}. 

\begin{table}[h]
 \centering
  \begin{tabular}{c|c|c}
   & \bb{stroke}	&	\bb{time interval} $[\tau_\text{cycle}]$\\
   \hline 
   (a) & right dissipative stroke& $\left[0,\frac{5}{24}\right) \cup \left[\frac{19}{24},1\right)$ \\
   (b) & isentropic stroke & $\left[\frac{5}{24},\frac{7}{24}\right)$ \\
   (c) & left dissipative stroke & $\left[\frac{7}{24},\frac{17}{24}\right)$ \\
   (d) & isentropic stroke & $\left[\frac{17}{24},\frac{19}{24}\right)$ \\
  \end{tabular}
  \caption{\label{table strokes} Division of the cycle into four strokes such that Eq.~(6) in the main text is 
  satisfied. }
\end{table}

%%%%%%%%%%%%%%%%%%%%%%%%%%%%%%%%%%%%%%%%%%%%%%%%%%%%%%%%%%%%%%%%%%%%%%%%%%%%%%%%%%%%%%%%%%%%%%%%%%%%%%%%%%%%%%%%%%%%%%%%
\section{Parameter choice for numerical simulations}
\label{sec parameter choice}

The electron shuttle can be realized using different experimental setups. We here focus on the case where the 
oscillatory degree of freedom is a nanopillar as sketched in Fig.~1 in the main text and realized in 
Refs.~\cite{KimQinBlickAPL2007, KimQinBlickNJP2010, KimParkBlickPRL2010, KimPradaBlickACS2012}. The material parameters typical for such experiments and used in this work are summarized below and in Table~\ref{tab:Parameters}. 

The bias voltage for electron shuttles can be tuned over a large regime. Here, we use a value of $V=25$V, which is a 
bit larger than experimentally reported values, but guarantees a clearly visible regime of self-oscillations and a 
simpler numerical treatment. We remark that the threshold value for the onset of self-oscillations is in our 
model around $V^* \approx 10$~V for the parameters chosen here. In reality, for $V=25$V we no longer expect the 
ultrastrong Coulomb blockade assumption to work well, which means that multiple electrons can hop on the dot. 
Importantly, this effect does not change the general conclusions reported in this paper and it can be easily accounted 
for (see the final conclusions in the main text). Furthermore, in all our calculations we choose a temperature of 
$T=1$K and set the chemical potentials as $\mu_L = \epsilon_0 + eV/2$ and $\mu_R = \epsilon_0 - eV/2$, which eliminates 
the dependence on the on-site energy $\epsilon_0$ in all equations. 

Since we consider transient dynamics, the choice of initial conditions and running time is important. Here, we choose 
the initial conditions $x_0 = 6.0$ nm, $v_0 = 0.0~\text{nm}/\text{ns}$ and $P_q(0) = \delta_{q,1}$ (filled dot). 
The simulation runs for $t_f = 250$~ns, which corresponds to roughly 150 cycles, and we average over 1000 trajectories.

\begin{table}[h]
\begin{tabular}{l|l|l|l}
\hline
\hline
Parameter & Value & Units & Source\\
\hline
 $\omega$ & $0.25$ & GHz & \cite{KimParkBlickPRL2010}\\ 
 $m$ & $20\times 10^{-19}$ & kg & \cite{KimParkBlickPRL2010, KimPradaBlickACS2012}\\
 $\lambda$ & $1$ & nm & \cite{KimPradaBlickACS2012}\\
 $\alpha$ & $0.01$ & $\text{nm}^{-1}$ & Estimated from \cite{KimQinBlickAPL2007, KimQinBlickNJP2010}\\
 $\gamma$ & $0.05\times 10^{-10}$ & $\text{kg}/\text{s}$ & \cite{KimPradaBlickACS2012}\\
 $\Gamma_0$ & $0.01$ & GHz & \cite{KimPradaBlickACS2012}\\
 $V$ & $25.0$ & V & \\
 $T$ & $1$ & K & \\
 \hline
 \hline
\end{tabular}
\label{tab:Parameters}
\caption{Parameters used in this work.}
\end{table}

%%%%%%%%%%%%%%%%%%%%%%%%%%%%%%%%%%%%%%%%%%%%%%%%%%%%%%%%%%%%%%%%%%%%%%%%%%%%%%%%%%%%%%%%%%%%%%%%%%%%%%%%%%%%%%%%%%%%%%%%
\section{Precise thermodynamic definitions for the reduced model}
\label{sec thermo reduced}

The internal energy and entropy of the dot are defined as follows: 
\begin{equation}\label{eq SM D def U and S}
 \C U_D(t) \equiv \sum_q [\epsilon_0-\alpha eVx(t)] q \C P_q(t), ~~~
 \C S_D(t) \equiv -k_B\sum_q \C P_q(t)\ln \C P_q(t).
\end{equation}
The heat flow and chemical work rate are composed out of the energy and matter fluxes as usual: 
$\dot{\C Q}_\nu(t) = \C J_\nu(t) - \mu_\nu\C I_\nu(t)$ and $\dot{\C W}_\text{chem}(t) = \sum_\nu\mu_\nu\C I_\nu(t)$. 
They are defined as 
\begin{equation}\label{eq SM D currents}
 \C I_\nu(t) \equiv \sum_{q,q'} q R_{q,q'}^\nu(x_t) \C P_{q'}(t), ~~~
 \C J_\nu(t) \equiv [\epsilon_0-\alpha eVx(t)] \C I_\nu(t).
\end{equation}

%%%%%%%%%%%%%%%%%%%%%%%%%%%%%%%%%%%%%%%%%%%%%%%%%%%%%%%%%%%%%%%%%%%%%%%%%%%%%%%%%%%%%%%%%%%%%%%%%%%%%%%%%%%%%%%%%%%%%%%%
\section{Cycle analysis of the reduced model in terms of thermodynamic strokes}
\label{sec cyle analysis reduced}

In this section we denote by $\tau_{a,b,c,d}$ the time-intervals defined in Table~\ref{table strokes} for brevity. 
Furthermore, for definiteness we focus on the analysis of the first cycle $[0,\tau_\text{cycle}]$. Extending our 
result below to further cycles is merely a matter of notation. 

Based on this, the state of the dot within the cycle analysis at time $t\in[0,\tau_\text{cycle}]$, denoted 
$[\C P_q(t)]_\text{cycle}$, can be written in the compact form 
\begin{equation}
 [\C P_q(t)]_\text{cycle} = 
 \C T_+ \exp\left\{\int_{\tau_a\cap[0,t]} dt R^L(x_t) + \int_{\tau_c\cap[0,t]} dt R^R(x_t)\right\} \C P_q(0),
\end{equation}
where $\C T_+$ denotes the time-ordering operator. If condition~(6) in the main text is satisfied, then 
we have 
\begin{equation}
 [\C P_q(t)]_\text{cycle} \approx \C P_q(t) = 
 \C T_+ \exp\left\{\int_0^t dt [R^L(x_t) + R^R(x_t)]\right\} \C P_q(0).
\end{equation}
The claim is now that this is sufficent to demonstrate that the thermodynamic analysis of the cycle coincides with 
the analysis of Sec.~\ref{sec thermo reduced}. 

To this end, we first note that the definition of the state functions internal energy and system entropy are the same 
as in Sec.~\ref{sec thermo reduced}, see Eq.~(\ref{eq SM D def U and S}), with $\C P_q(t)$ replaced by 
$[\C P_q(t)]_\text{cycle}$. Thus, clearly, if $\C P_q(t) \approx [\C P_q(t)]_\text{cycle}$, then 
\begin{equation}
 \C U_D \approx [\C U_D]_\text{cycle} ~~~ \text{and} ~~~ \C S_D \approx [\C S_D]_\text{cycle},
\end{equation}
where we used $[\C X]_\text{cycle}$ to denote a thermodynamic quantity $\C X$ in our cycle analysis. 
Furthermore, the definition of mechanical work during stroke $s$ ($s\in\{a,b,c,d\}$) is 
\begin{equation}
 [\C W_{\rm mech}^{(s)}]_\text{cycle} 
 = \int_{\tau_s} dt [\C P_1(t)]_\text{cycle} \frac{\partial\epsilon(x_t)}{\partial t}.
\end{equation}
Again, if $\C P_q(t) \approx [\C P_q(t)]_\text{cycle}$, we clearly have 
\begin{equation}
 \sum_s [\C W_{\rm mech}^{(s)}]_\text{cycle} \approx 
 \C W_\text{mech} = \int_0^{\tau_\text{cycle}} dt \C P_1(t) \frac{\partial\epsilon(x_t)}{\partial t}.
\end{equation}

We now continue with a step-by-step analysis of the thermodynamic cyle

\begin{enumerate}
 \item[(a)] \underline{Left dissipative stroke:} 
 The dot is only coupled to the left lead and the first and second law read 
 \begin{equation}
  \Delta\C U_S = [\C Q_L]_\text{cycle} + [\C W_\text{chem}]_\text{cycle} + \C W_\text{mech}, 
  ~~~ \Delta S_D - [\C Q_L]_\text{cycle}/T \ge 0.
 \end{equation}
 The not yet defined quantities appearing here are 
 \begin{equation}
  [\C Q_L]_\text{cycle} = \int_{\tau_a} dt [\C J_L(t) - \mu_L \C I_L(t)], 
  ~~~ [\C W_\text{chem}]_\text{cycle} = \int_{\tau_a} dt \mu_L I_L(t)
 \end{equation}
 with the energy and matter current defined in Eq.~(\ref{eq SM D currents}). 
 \item[(b)] \underline{Isentropic stroke:} One has 
 \begin{equation}
  \Delta\C U_S = \C W_\text{mech}, ~~~ [\C Q_L]_\text{cycle} = [\C Q_R]_\text{cycle} = 0, 
  ~~~ [\C W_\text{chem}]_\text{cycle} = 0, ~~~ \Delta S_D = 0.
 \end{equation}
 \item[(c)] \underline{Left dissipative stroke:} 
 Everything as in (a) with $L$ replaced by $R$ and $\tau_a$ replaced by $\tau_c$. 
 \item[(d)] \underline{Isentropic stroke:} Identical to (b). 
\end{enumerate}

Thus, to finally guarantee that the cycle analysis matches the analysis from Sec.~(D), we recall that the bare 
tunneling rates $\Gamma_{L,R}(x_t)$ are contained in the rate matrix $R^{L,R}(x_t)$ as an overall factor. Thus, 
\emph{if} condition~(6) in the main text is satisfied, we observe that 
\begin{align}
 \C Q_L &= \int_{[0,\tau_\text{cycle}]} dt [\C J_L(t) - \mu_L\C I_L(t)] 
 \approx \int_{\tau_a} dt [\C J_L(t) - \mu_L\C I_L(t)] = [\C Q_L]_\text{cycle}, \\
 \C Q_R &= \int_{[0,\tau_\text{cycle}]} dt [\C J_R(t) - \mu_R\C I_R(t)] 
 \approx \int_{\tau_c} dt [\C J_R(t) - \mu_R\C I_R(t)] = [\C Q_R]_\text{cycle}.
\end{align}
This shows that the thermodynamic analysis in terms of a cycle is \emph{automatically} consistent if the analysis 
of Sec.~\ref{sec thermo reduced} is consistent (which requires large mass $m$ and small friction $\gamma$) 
\emph{and} condition~(6) in the main text is satisified. 

\end{document}